\documentclass[prd,twocolumn,preprintnumbers,superscriptaddress]{revtex4-2}
\usepackage{amsmath}
\usepackage{color}
\usepackage{ulem}
\usepackage{graphicx}
\usepackage{hyperref}
\usepackage{blindtext}
\usepackage{amsfonts}
\hypersetup{colorlinks, citecolor=blue, linkcolor=blue, urlcolor=magenta}
\usepackage[caption=false]{subfig}

\usepackage[american]{babel}
\usepackage{microtype}
 \usepackage[table,xcdraw]{xcolor}
\definecolor{darkblue}{rgb}{0.0,0.0,0.6}

\newcommand{\Mpl}{M_{\rm Pl}}

\usepackage{geometry}
\usepackage{fullpage}
\begin{document}

\title{ Is the NANOGrav detection evidence of resonant particle creation during inflation?}

\author{M. R. Gangopadhyay}
\email{mayukh$\_$ccsp@sgtuniversity.org}
\affiliation{Centre For Cosmology and Science Popularization (CCSP), SGT University, Gurugram, Delhi- NCR, Haryana- 122505, India.}
\author{V. V. Godithi}
\email{ms21203@iisermohali.ac.in}
\affiliation{Department of Physics, IISER Mohali,  S.A.S Nagar, Mohali, Punjab- 140306}
\affiliation{Centre For Cosmology and Science Popularization (CCSP), SGT University, Gurugram, Delhi- NCR, Haryana- 122505, India.}
\author{K. Ichiki}
\email{ichiki.kiyotomo.a9@f.mail.nagoya-u.ac.jp}
\affiliation{Department of Physics and Astrophysics, Nagoya University, Chikusa-ku, Nagoya, 464-8602, Japan}
\author{R. Inui}
\email{inui.ryoto.a3@s.mail.nagoya-u.ac.jp}
\affiliation{Department of Physics and Astrophysics, Nagoya University, Chikusa-ku, Nagoya, 464-8602, Japan}
\author{T. Kajino}
\email{kajino@buaa.edu.cn}
\affiliation{School of Physics, and International Research Center for Big-Bang Cosmology and Element Genesis, \\
Beihang University, Beijing 100183, China}
\affiliation{Division of Science, National Astronomical Observatory of Japan, \\
2-21-1 Osawa, Mitaka, Tokyo 181-8588, Japan}
\affiliation{Graduate School of Science, The University of Tokyo, \\
7-3-1 Hongo, Bunkyo-ku, Tokyo 113-033, Japan}

\author{A. Manusankar}
\email{manusankaranand@gmail.com}
\affiliation{Department of Physics,Cochin University of Science  and Technology,Kalamassery,Kochi,Kerala-682022}
\affiliation{Centre For Cosmology and Science Popularization (CCSP), SGT University, Gurugram, Delhi- NCR, Haryana- 122505, India.}
\author{G. J. Mathews}
\email{gmathews@nd.edu}
\affiliation{Center for Astrophysics,
Department of Physics and Astronomy, University of Notre Dame, Notre Dame, IN 46556, USA.}
\author{Yogesh}
\email{yogeshjjmi@gmail.com }
\affiliation{Centre For Cosmology and Science Popularization (CCSP), SGT University, Gurugram, Delhi- NCR, Haryana- 122505, India.}

\begin{abstract}
We show that the recently reported cosmic gravitational wave background by the NANOGrav 15-year collaboration may be the result of resonant particle creation during inflation.  For the appropriate amplitude and particle mass an enhancement of the primordial scalar power spectrum could induce  Secondary Induced Gravitational Waves (SIGW) which will appear on a scale corresponding to the frequency of the NANOGrav detection. Since the resonant creation will have an effect comparable to that of a delta function increment as studied by the NANOGrav 15-year collaboration, our study indicates that the low-frequency Pulsar Timing Array (PTA) data could reveal the aspects of the physics during inflation through the detection of a cosmic background of Gravitational Waves (GW). 
\end{abstract}
\maketitle
\section{Introduction}
Evidence for the existence of a gravitational-wave background has recently been reported \cite{NANOGrav:2023hde} in the $n$Hz frequency range. The latest announcement from various Pulsar Timing Array (PTA)
collaborations, NANOGrav, EPTA, PPTA, and CPTA \cite{NANOGrav:2023hde}-\cite{Xu:2023wog} has created a new wave of motivation to seek a cosmological explanation of such an excess in the gravitational wave background as observed.

Although the observations have been interpreted as a stochastic background possibly due to binary mergers of supermassive black holes, it is worthwhile to explore other possible cosmological sources of this gravitational wave background.  Indeed, a number of works have explored possible origins of a stochastic gravitational wave background  \cite{Vaskonen21,Kohri21,Correa23, Deluca21, Domenech22,Inomata21,Das:2023nmm,Talebian23,Nakai21,Ratzinger21,Arzoumanian21,Bringmann23,Blasi21,Ellis21, Ellis23, Wang23, Higaki16,Ananda07, Bauman07, Bugaev10,Assadullahi09,Alabidi12,Cai19,Mansoori23,Ghosh:2023aum,Kawai:2023nqs,Ragavendra:2020sop,Ragavendra:2020vud,Ragavendra:2023ret}. In particular, we recently reported \cite{Correa23} upon the gravitational wave background generated by the induced spectrum of tensor fluctuation during warm natural inflation.  However, the magnitude of the power spectrum was significantly below that required to explain the recently reported gravitational-wave background.  Some enhancement in the primordial power spectrum is required on the frequency scale of the NANOGrav observation. Recently, considerable effort has been made to explain an excess of tensor power in the $n$Hz frequency range [cf. \cite{Choudhury:2023wrm}-\cite{Li:2023xtl}].

For example, in Ref.~\cite{Mansoori23} a chaotic inflation model was proposed with $\mathbb{T}^2$ inflation, i.e. that incorporates terms proportional to $T_{\mu \nu}^{\mu \nu}$.  In this scenario, parameters can be tuned to enhance the power spectrum at the appropriate scale.  In this paper, however, we note another possibility to generate an enhancement (or dip) in the primordial power spectrum, i.e. resonant particle creation during inflation as described in a number of works \cite{Chung00, Mathews04, Mathews15, Gangopadhyay18}.  In this work, we explore the possibility of producing the desired enhancement in the NanoGRAV frequency range via this mechanism.
\section{Resonant Particle Production During Inflation}
\label{resonance creation}
In the standard inflationary scenario, the inflaton field does not interact with any other fields during the inflation. In a minimal extension, if the inflaton is allowed to couple to at least one massive particle with mass of order of the inflaton field value \cite{Chung00}, then a resonant production of such particles  will  take place as the inflaton field reaches a critical value. Thus, due to this resonant production, new features can appear in the primordial power spectrum. Depending on the scale at which the resonant particle production will take place and the sign of the interaction, this can be associated with an excess in the region of the primordial power spectrum associated with the scale of the resonance.

In the slow-roll approximation, 
The amplitude of the density fluctuation when it crosses the Hubble
radius $\delta_H(k)$ is given by:
\begin{equation}
\delta_H(k) \approx {H^2
\over 5 \pi \dot \phi}~~,
\label{pert}
\end{equation}
where $H$ is the Hubble parameter and  $\dot \phi$ is 
the inflaton field time derivative when the comoving wavenumber $k$ crosses the
Hubble radius during inflation. 

In the case of  resonant particle production,
the inflaton field will lose energy as it approaches the critical field value where the resonance will take place. Thus, the conjugate momentum of the inflation field $\dot \phi$ will decrease.  Then there will be an enhancement in 
$\delta_H(k)$ (primordial power spectrum) for those wave numbers which
exit the horizon around the scale of the resonant particle production epoch.

If the inflaton field has a simple Yukawa coupling to a fermion field $\psi$ of mass $m$, then the interaction Lagrangian density is given by:
\begin{equation}
{\cal L}_Y = -\lambda \phi \bar \psi \psi~~.
\label{eq:yukawacoupling}
\end{equation}
In this case the equation of motion
for the inflaton field can be written as:
\begin{equation}
\ddot \phi + 3H \dot \phi + {dV \over d\phi} - 
N \lambda  \langle \bar \psi \psi \rangle = 0 ~~,
\label{eom}
\end{equation}
for $N$ fermions of mass $m$ coupled to the inflaton.

The effective mass of the fermion is then given by:
\begin{equation}
M(\phi) = m - \lambda \phi ~~. 
\end{equation}
The effective mass term thus vanishes as the inflaton field value reaches  $\phi_* = m/\lambda$. Hence, a resonant creation of this fermion field will take place as $\phi \rightarrow  \phi_*$. 

Following \cite{Chung00,Mathews04,Mathews15,Gangopadhyay18}, the perturbation spectrum can be written as:
\begin{equation}
\label{eq:deltaanalytic}
\delta_H(k) =  \frac{ [\delta_H(k)]_{\lambda=0}}
{1-\theta(a-a_*) |\dot{\phi}_*|^{-1}
N\lambda n_* H_*^{-1}(a_*/a)^3\ln(a/a_*)}~~ .
\end{equation}
The perturbation spectrum Eq.~(\ref{eq:deltaanalytic}) 
can then be reduced \cite{Chung00} to a simple  two-parameter function.
\begin{equation}
\label{eq:fit}
\delta_H(k) = \frac{\left[\delta_H(k)\right]_{\lambda=0}}
{1-\theta(k-k_*) A(k_*/k)^3\ln(k/k_*)}~~~,
\end{equation}
where the coefficient $A$ and characteristic wave number $k_*$ ($k/k_*
\ge 1$)  determine the amplitude of the enhancement and the scale of the enhancement, respectively of the scalar power spectrum. 

The coefficient $A$
can be related directly to the coupling constant $\lambda$ 
using the approximation
\cite{Chung00,Mathews04,Mathews15, Gangopadhyay18} for the
particle production Bogoliubov coefficient 
\begin{equation}
|\beta_k|^2 = \exp\left( \frac{-\pi k^2}{a_*^2  \lambda |\dot \phi_*| }\right)~~.
\end{equation}

Then,
\begin{equation}
\label{eq:nstar}
n_* = \frac{2}{\pi^2}\int_0^\infty dk_p \, k_p^2 \, 
|\beta_k|^2 =
\frac{ \lambda^{3/2}}{2\pi^3}
|\dot{\phi}_*|^{3/2}~~ .
\end{equation}
This gives:
\begin{eqnarray}
A & = & \frac{ N \lambda^{5/2}}{2 \pi^3}
\frac{\sqrt{|\dot{\phi}_*|}}{H_*}\\
& \approx & \frac{N \lambda^{5/2}}{2 \sqrt{5} \pi^{7/2}}
\frac{1}{\sqrt{\delta_H(k_*)|_{\lambda=0}}}~~.
\label{eq:alamrelation}
\end{eqnarray}
where we have used the usual approximation for the primordial slow
roll inflationary spectrum \cite{Liddle,cmbinflate}.  This means that
regardless of the exact nature of the inflationary scenario, for any
fixed inflationary spectrum $\delta_H(k)|_{\lambda=0}$ without the
back reaction, resonant particle production is possible and can lead to an excess in the amplitude of the density perturbation.  Given that
the CMB normalization requires $\delta_H(k)|_{\lambda=0}\sim
10^{-5}$, we then have 
\begin{equation}
A \sim 1.3 N  \lambda^{5/2}.
\label{norm}
\end{equation}

From Eq.~(\ref{norm}),   one can deduce that $\lambda \le 1$ requires $N>1$ as expected for the given values of $A$.

\begin{figure}[h]
    \centering
    \includegraphics[height=3.0in,width=3.5in]{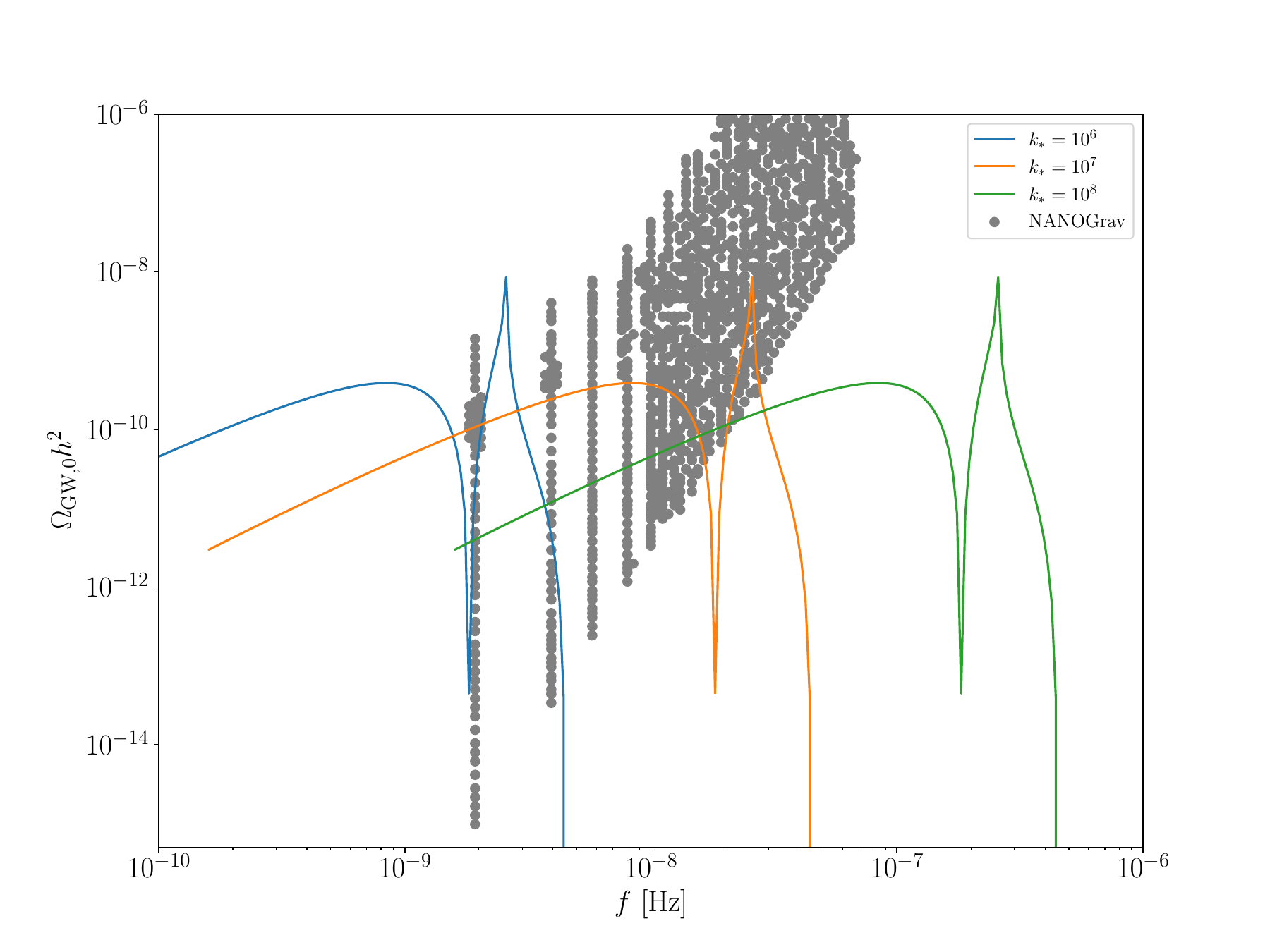}
    \caption{Closure contribution from SIGW for three different values of the wave number with $A = 8.15457$, $\alpha=0$, and $\beta =0$. The grey dotted points represent the NANOGrav-15 data.  }
    \label{fig:Omggw_alpha0}
\end{figure}

\begin{figure}[h]
    \centering
    \includegraphics[height=3.0in,width=3.5in]{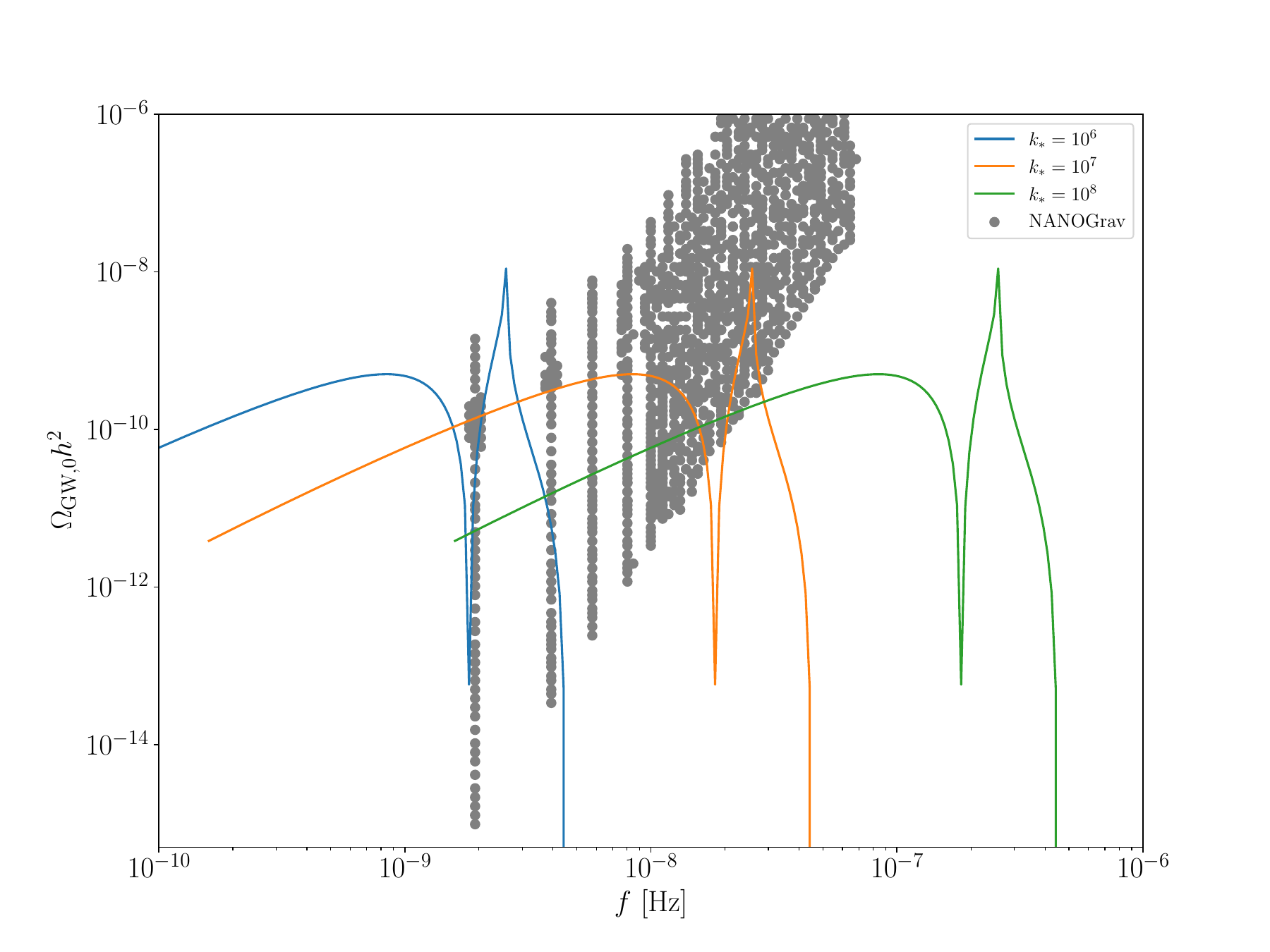}
    \caption{Closure contribution from SIGW for three different values of the wave number with $A = 8.15452$, $\alpha=0.002$ and $\beta =0$. The grey dotted points represent the NANOGrav-15 data.  }
    \label{fig:Omggw_alpha0002}
\end{figure}

\noindent
Generic slow-roll inflation models can produce  perturbations which are generally well modeled by the following adiabatic scalar and tensor components\cite{pl2}:
\begin{align}
\label{eq:powerlaw} \ln \mathcal{P_R}(k) =& \ln \mathcal{P}_0(k)
+ \frac{1}{2} \frac{d \ln n_\mathrm{s}}{d \ln k} \ln(k/k_*)^2\nonumber\\
&\quad + \frac{1}{6} \frac{d^2 \ln n_\mathrm{s}}{d \ln k^2} \ln(k/k_*)^3  + \ldots ,\\
\ln \mathcal{P}_\mathrm{t}(k) =& \ln (r A_\mathrm{s}) + n_\mathrm{t} \ln(k/k_*) + \ldots  ,
\end{align}
Allowing the contribution from the scale dependence of the scalar spectral index, $n_\mathrm{s}$,  
modelled by a running $\alpha(={d \ln n_\mathrm{s}}/{d \ln k})$,  and running 
of the running, $\beta(={d^2 \ln n_\mathrm{s}}/{d (\ln k)^2})$.  For the scalar perturbation, Eq. (\ref{eq:powerlaw}), can be re-written in a more familiar form:
\begin{equation}
\mathcal{P}_{\mathcal{R}}(k)=A_{s}\left( \frac{ k}{k_{p}}\right)^{n_{s}-1+\frac{\alpha_{s}}{2}  \ln\left(  \frac{k}{k_{p}}\right) +\frac{\beta_{s}}{6}  \left(\ln\left(  \frac{k}{k_{p}}\right)\right)^{2}}
\label{eq:PS}
\end{equation}
where $k_p$, is the Planck pivot scale,  we take $k_{p}=0.05$.  For our calculation, the Planck 2018 TT(TT,TE,EE)+lowE+lensing qouted central values are used for $A_s$,  $n_s$, $\alpha$ and $\beta$ as follows:
\begin{align}
\ln(10^{10} A_\mathrm{s})&= 3.044 \pm 0.014 \\
n_\mathrm{s} &= 0.9587 \pm 0.0056  \,\,(0.9625 \pm 0.0048) \,, \\
d n_\mathrm{s}/d \ln k &= 0.013 \pm 0.012 \,\,(0.002 \pm 0.010) \,, \\
d^2 n_\mathrm{s}/d \ln k^2 &= 0.022 \pm 0.012 \,\,(0.010 \pm 0.013) \,,
\end{align}
For our model, the numerator of  Eq.~(\ref{eq:fit}) is related to Eq.~(\ref{eq:PS}) ($(\delta_H(k)|_{\lambda=0})^2 =\mathcal{P}_{\mathcal{R}}(k) $). After plugging that into Eq.~(\ref{eq:fit}), we calculate the GW spectrum keeping $\beta=~0$ and using the central values of $n_s$ and $\ln(10^{10} A_\mathrm{s})$. We adopt two different values for $\alpha=~0.00$ and $\alpha=~0.002$, (the second value corresponds to the central value as quoted in \cite{pl2}). The corresponding GW spectra are shown in Figs. \ref{fig:Omggw_alpha0} and \ref{fig:Omggw_alpha0002}.
\section{Secondary Induced Gravity Wave Production}
\label{analysis}

It has been shown in the literature that the formation of PBHs\cite{Zeldovich:1967lct}-\cite{Choudhury:2013woa} and scalar-induced gravitational waves are inherently connected with each other\cite{Saito,Garcia-Bellido:2017aan,Inomata:2016rbd,Inomata:2018epa}. Due to the nonlinear nature of gravity, the same scalar perturbation that dictates the formation of the PBHs can induce the production of GWs when the relevant scale enters the horizon \cite{Bauman07}(see Ref. \cite{Yuan:2021qgz,Domenech:2021ztg} for review). A metric that contains both the scalar and tensor perturbation can be used to model the GWs \cite{Kohri:2018awv,Cappelluti22,Naidu22}. In the conformal Newtonian gauge, the metric is defined as: 
\begin{eqnarray}
    ds^2 &=& a(\eta)^2\biggl[ -\left(1+2 \Phi\right)d\eta^2 \nonumber \\
    &+&\left(\left(1-2 \Psi\right)\delta_{ij}+ \frac{h_{ij}}{2}\right)dx^i dx^j
\biggr] ~~ ,
\end{eqnarray}
where $\eta$, the conformal time, $a$ is the scale factor, $\Psi $ and $\Phi$ are the scalar perturbations, while  $h_{ij}$ is the tensor perturbation. While neglecting the first-order GWs, anisotropic stress (since the contribution is small), and vector perturbations, and assuming  $\Psi =\Phi$, we write the second-order graviton action as:
\begin{eqnarray}
    S=\frac{\Mpl^2}{32}\int d\eta~ d^3x~ a^2 \left[\left(h'_{ij}\right)^2-\left(\partial_l h_{ij}\right)^2\right]\, ,
    \label{acttion:graviton}
\end{eqnarray}
where, $M_{pl}$ is the Planck mass and ($'$) stands for the differentiation w.r.t to $\eta$. The $\partial_l h_{ij}$ signify the derivative of $h_{ij}$ w.r.t spatial coordinate $l$. The tensor modes can be decomposed into Fourier components and written in the usual form
\begin{eqnarray}
    h_{ij}(\eta, {\bf{x}})=\int \frac{d^3 ~k}{(2\pi)^{3/2}} \sum_{s=+, \times}\epsilon_{ij}^{s}(k)h^{s}_{{\bf k}}(\eta) e^{i{\bf k x}}~~,
\end{eqnarray}
where $\epsilon^{+}_{ij}(k)$ and $\epsilon^{\times}_{ij}(k)$ are the transverse traceless polarization vectors. In terms of  the normalized orthogonal vectors $e_i({\bf k}), \bar{e}_i({\bf k})$ the polarization vectors can be written:

\begin{eqnarray}
   \epsilon_{ij}^{+}(\bf k)&=&\frac{1}{\sqrt{2}}\left[ e_i({\bf k}) e_j({\bf k})-\bar{e}_i({\bf k}) \bar{e}_j({\bf k})\right] \\
   \epsilon_{ij}^{\times}(\bf k)&=&\frac{1}{\sqrt{2}}\left[ e_i({\bf k}) \bar{e}_j({\bf k})+\bar{e}_i({\bf k}) e_j({\bf k})\right]~~.
\end{eqnarray}
The dimensionless power can be written as: 
\begin{eqnarray}
  \bigl\langle h_{\bf k}^\lambda (\eta) h_{\bf k'}^{\lambda'}(\eta)  \bigr\rangle = \delta_{\lambda \lambda'}\frac{2 \pi^2}{k^3} \delta(k+k') \mathcal{P}_{h}(k, \eta)\, ,
  \label{eq:Ten_PS}
\end{eqnarray}
where the polarization indices are represented by  $\lambda, \lambda'= \{+, \times\}$. Following \cite{Cappelluti22} one can obtain the tensor equation of motion induced by the scalar fluctuations $\Phi$,
\begin{align}
h''_{\bf k}(\eta) + 2 \mathcal{H} h'_{\bf k}(\eta)+ k^2 h_{\bf k}(\eta) =& 4 S_{\bf k}(\eta)~~.  \label{EOM_h} 
\end{align}
Defining $\mathcal{H}\equiv {a'}/{a}= a H $, the source term $S_{\bf k}$ becomes:
\begin{eqnarray}
S_{\bf k} &=& \int \frac{\text{d}^3 q}{(2 \pi)^{3/2}} e_{ij}({\bf k}) q_i q_j \biggl( 2\Phi_{\bf q}  \Phi_{{\bf k}-{\bf q}} +  \\
&& \frac{4}{3(1+w)} \left( \mathcal{H}^{-1} \Phi'_{\bf q} + \Phi_{\bf q}\right) \left( \mathcal{H}^{-1} \Phi'_{{\bf k}-{\bf q}} + \Phi_{{\bf k}-{\bf q}} \right)  \biggr) ~,\nonumber
\end{eqnarray}
By making the use of standard relation
$-2 \dot{H} = \rho + P =( 1+w) \rho = 3 (1+w)H^2$, where $w = P/\rho$ is equation-of-state parameter. The Fourier modes of the gravitational potential $\Phi_{\bf k}$ are comparable to those of the tensor mode. Employing the Green's Function method to evaluate the tensor mode $h_{\bf k}(\eta)$, one obtains:
\begin{align}
a(\eta) h_{\bf k}(\eta) = 4 \int^\eta \text{d}\bar{\eta} G_{\bf k}(\eta, \bar{\eta}) a(\bar{\eta}) S_{\bf k}(\bar{\eta})~~.
\label{eq:Ten_fluc}
\end{align}
where, the Green's function $G_{\bf k}(\eta, \bar{\eta})$ is determined from:
\begin{align}
G_{\bf k}''(\eta, \bar{\eta}) +\left( k^2 - \frac{ a''(\eta)}{a(\eta)}\right) G_{\bf k}(\eta, \bar{\eta}) = \delta (\eta - \bar{\eta}), \label{EOM_Green}
\end{align}
and derivatives are with respect to $\eta$.
Furthermore the equation of motion of the gravitational potential $\Phi_{\bf k}$ can be written as (e.g.~\cite{Mukhanov05}):
\begin{eqnarray}
&\Phi''_{\bf k}& + 3 \mathcal{H} (1 + c_{\text{s}}^2) \Phi'_{\bf k} + (2 \mathcal{H}'+(1+3 c_{\text{s}}^2)\mathcal{H}^2 +c_{\text{s}}^2 k^2) \Phi_{\bf k} 
\nonumber \\
&=& \frac{a^2}{2} \tau \delta S,  \label{EOM_Phi_complete}
\end{eqnarray}
where in the above expression, the sound speed $ c_{\text{s}}^2 = w$ and temperature  $\tau$ are represented by $\delta P = c_{\text{s}}^2 \delta \rho + \tau \delta S$, where  $S$ is the entropy density. When the entropy perturbations are absent, the above equation simplifies to:
\begin{align}
\Phi''_{\bf k}(\eta) + \frac{6(1+w)}{(1+3w)\eta } \Phi'_{\bf k}(\eta) + w k^2 \Phi_{\bf k}(\eta)=0~~. \label{EOM_Phi}
\end{align}
From the relation $\Phi_{\bf k} = \Phi(k \eta) \phi_{\bf k}$ the primordial value $\phi_{\bf k}$ can be derived.  In addition, the fact that  the transfer function becomes unity well before the horizon entry leads to the following relation between the primordial value and curvature perturbation:
\begin{align}
\langle \phi_{\bf k} \phi_{\bf k'} \rangle = \delta ({\bf k}+{\bf k}' ) \frac{2\pi^2}{k^3} \left( \frac{3+3w}{5+3w} \right)^2 \mathcal{P}_\zeta (k)~~,
\label{pzeta}
\end{align}
where, the equation of state parameter $\omega$ needs to be evaluated well before the horizon entry. Similarly, the primordial value $\phi_{\bf k}$ needs to be evaluated before the horizon entry. 

If one neglects the non-Gaussanity of the primordial curvature perturbations, the correlation function  $\langle S_{\bf k} (\eta) S_{\bf k'}(\eta') \rangle$ can be obtained. Furthermore, by comparing $\langle S_{\bf k} (\eta) S_{\bf k'}(\eta') \rangle$ with  $\mathcal{P}_h$, using Eqs. (\ref{eq:Ten_PS}) and (\ref{eq:Ten_fluc}) and doing some algebraic simplification,  one can  extract the power spectrum from the curvature perturbation $P_\zeta$  \cite{Kohri:2018awv}:
\begin{eqnarray}
\mathcal{P}_h (\eta, k) &=&  4  
 \int_0^\infty \text{d}v \int_{\left| 1-v \right |}^{1+v}\text{d} u \left( \frac{4v^2 - (1+v^2-u^2)^2}{4vu} \right)^2 \nonumber \\
 &\times &I^2 (v,u,x) \mathcal{P}_\zeta ( k v ) \mathcal{P}_\zeta ( k u ), \label{P_h}
\end{eqnarray}
where $x\equiv k \eta$, and
\begin{align}
I(v,u,x)= \int_0^x \text{d}\bar{x} \frac{a (\bar{\eta})}{a(\eta)} k G_k (\eta, \bar{\eta}) f (v,u,\bar{x}),   \label{I}
\end{align}
with
\begin{eqnarray}
f (v ,u ,\bar{x}) &= & \frac{6(w+1)}{3w+5}\Phi(v\bar{x})\Phi(u\bar{x}) + \frac{6(1+3w)(w+1)}{(3w+5)^2} \nonumber \\
&\times &\left( \bar{x}\partial_{\bar{\eta}}\Phi(v\bar{x})\Phi(u\bar{x}) +\bar{x}\partial_{\bar{\eta}} \Phi(u\bar{x})\Phi(v\bar{x}) \right) \nonumber \\
& +& \frac{3(1+3w)^2(1+w)}{(3w+5)^2}\times \bar{x}^2 \partial_{\bar{\eta}}\Phi(v\bar{x})\partial_{\bar{\eta}}\Phi(u\bar{x}),\nonumber\\
\end{eqnarray}
and $\bar{x} \equiv  k \bar{\eta}$, while $\mathcal{H}=aH=2/[(1+3w)\eta]$. The quantity $f(u, v, \bar{x})$ contains the information of the source term. By making the following change of  variables $u+v-1\rightarrow t$ and $u-v \rightarrow s$, one can rewrite the integral \eqref{P_h} as 
\begin{eqnarray}
\mathcal{P}_h (\eta, k) &=&  2  
 \int_0^\infty \text{d}t \int_{ -1}^{1}\text{d} s \left( \frac{t(2+t)(s^2-1)}{(1-s+t)(1+s+t)} \right)^2 \nonumber \\
 &\times &I^2 (s,t,x) \mathcal{P}_\zeta ( k v ) \mathcal{P}_\zeta ( k u ) \label{P_h_st} ~~.
\end{eqnarray}
\begin{figure}[h]
    \centering
    \includegraphics[height=3.0in,width=3.5in]{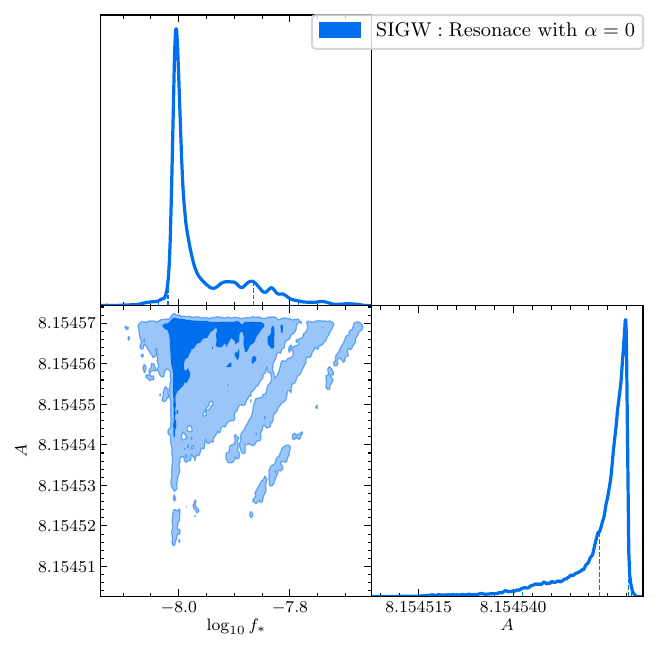}
    \caption{The current NANOGrav constraint on $A$ for  $\alpha=0, \beta =0$. }
    \label{fig:corner_alpha0}
\end{figure}
\begin{figure}[h]
    \centering
    \includegraphics[height=3.0in,width=3.5in]{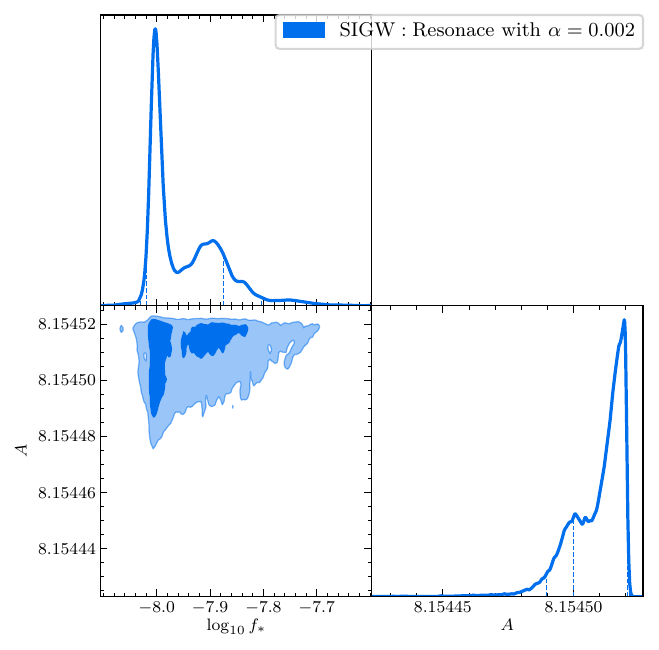}
    \caption{The current NANOGrav constraint on $A$ for  $\alpha=0.002, \beta =0$ case.  }
    \label{fig:corner_alpha0002}
\end{figure}
Following \cite{Kohri:2018awv}, in a radiation dominated universe one can compute the GWs by taking the limit ($ x\to\infty$) and taking the average of the oscillation  $I^2(s,t,x)$  one can write, 
\begin{multline}
     \overline{I^2(s,t,x\to\infty)} = \frac{288(-5+s^2+t(2+t))^2}{x^2(1-s+t)^6(1+s+t)^6} \\
     \times \Biggl[\frac{\pi^2}{4}(-5+s^2+t(2+t))^2 \theta(t-(\sqrt{3}-1))  \\
      +\Bigl( ~-(t-s+1)(t+s+1)  \\
      +\frac{1}{2}(-5+s^2+t(2+t)) \log{\left|\frac{-2+t(2+t)}{3-s^2}\right|} ~\Bigl) ~\Biggl] ~~, \label{avg_I2}
\end{multline}
where  an average over the oscillations is represented by the overline, and $\theta(x)$ represents the Heaviside step function. 

Using Eqs.~\eqref{P_h_st} and \eqref{avg_I2} we compute the power spectrum by taking advantage of another change of variable ($t+1\rightarrow \sqrt{r}$),
\begin{multline}
    \overline{\mathcal{P}_h(\eta, k)} =  \int_1^\infty \text{d}r \int_{ -1}^{1}\text{d} s \\
    \frac{72~(r-1)^2 \left(s^2-1\right)^2  \left(r+s^2-6\right)^2}{x^2 ~\sqrt{r} ~ \left(r-s^2\right)^8} \\
    \times \Biggl[  ~ \left(~ \left(r+s^2-6\right) ~\log \left(\left| \frac{3-r}{s^2-3}\right| \right)- 2 \left(r-s^2\right) ~ \right)^2 \\
    + \pi ^2 ~\left(r+s^2-6\right)^2 ~ \theta \left(\sqrt{r}-\sqrt{3}\right) ~ \Biggl] \\
    \times
    \mathcal{P}_\zeta\left(\frac{1}{2} k \left(\sqrt{r}-s\right)\right) 
    \mathcal{P}_\zeta\left(\frac{1}{2} k \left(\sqrt{r} + s\right)\right)~~.
\end{multline}
The above integral can be evaluated numerically
 using $\mathcal{P}_\zeta$ from Eq.~(\ref{pzeta}).
 
 In the subhorizon limit the GW energy density $\rho_{\text{GW}}(\eta)=\int \text{d}\ln k \rho_{\text{GW}}(\eta, k)$ can be computed as: 
 \begin{align}
\rho_{\text{GW}}=\frac{\Mpl^2}{16a^2} \left \langle \overline{h_{ij,k}h_{ij,k}} \right \rangle ,  \label{rho_GW}
\end{align}
 The parity invariance of the polarization modes is in such a manner that each mode gives an equal contribution to the energy density. The dimensionless closure parameter can then be used to describe the contribution from  the stochastic background of gravitational waves (GWs).
\begin{eqnarray}
\Omega_{\text{GW}}(\eta, k)&=& \frac{1}{\rho_{tot} (\eta)} \frac{d\rho_{\rm GW}(\eta, k) }{d\ln{k}}  \\
&= &\frac{\rho_{\text{GW}}(\eta,k)}{\rho_{\text{tot}}(\eta)}= \frac{1}{24} \left( \frac{k}{a(\eta)H(\eta)} \right)^2 \overline{\mathcal{P}_h(\eta, k)}~~, \nonumber
\label{Omega_GW}
\end{eqnarray}
For a radiation dominated universe we set $\left( \frac{k}{a(\eta)H(\eta)} \right)^2= k^2\eta^2=x^2$,  and following following \cite{Kohri:2018awv},   we obtained the present GW spectrum as:
\begin{align}
\Omega_{\text{GW}, 0}(k)= \frac{g_{*, rad}}{g_{*, 0}} \left( \frac{g_{*S, 0}}{g_{*S, rad}} \right)^{4/3} \Omega_{r,0}h^2 ~ \Omega_{\text{GW}}(\eta_c, k)~,
\label{Omega_GW_0}
\end{align}
where the subscript ($rad$) represents quantities computed when the perturbation is within the horizon during the radiation-dominated era so that  $\rho_{\text{GW} }$ is a constant percentage of radiation energy density. Additionally, we take $ \frac{g_{*, rad}}{g_{*, 0}} \left( \frac{g_{*S, 0}}{g_{*S, rad}} \right)^{4/3} 
\Omega_{r,0} h^2~=~6.4\times10^{-5}$. 
 Finally, we use the further relation between wave-number $k$ and gravitational wave frequency $f$ as:
\begin{align}
f=\frac{k}{2 \pi}=1.5\times10^{-15} \left( \frac{k}{1~\text{Mpc}^{-1}} \right) \text{Hz}~~.
\label{freq}
\end{align}

\begin{figure}[h]
    \centering
    \includegraphics[height=3.0in,width=3.5in]{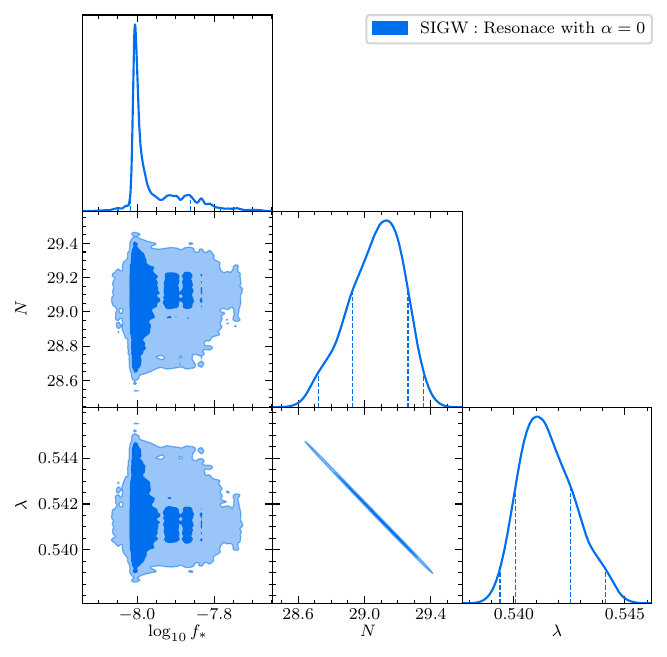}
    \caption{The current NANOGrav constraint on $N$ and $\lambda$ for $\alpha=0, \beta =0$ case.}
    \label{fig:corner_N_lambda_alpha0}
\end{figure}

\begin{figure}[h]
    \centering
    \includegraphics[height=3.0in,width=3.5in]{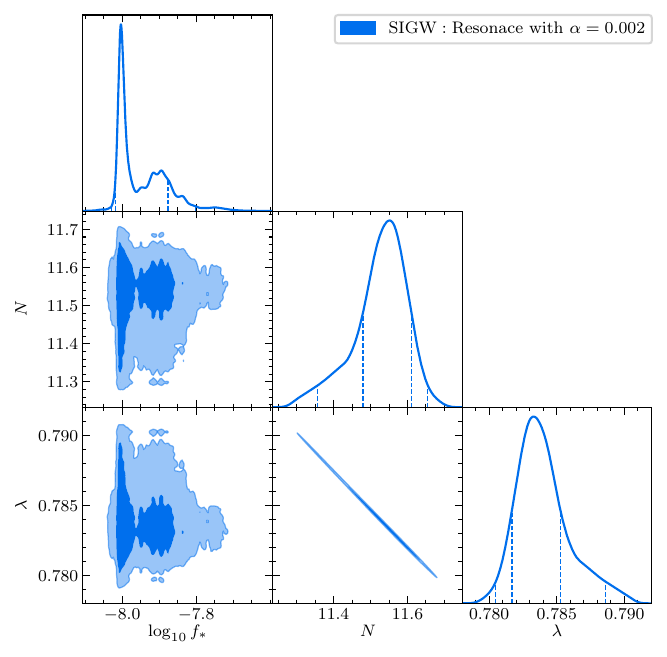}
    \caption{The current NANOGrav constraint on $N$ and $\lambda$ for $\alpha=0.002, \beta =0$ case.}
    \label{fig:corner_N_lambda_alpha0002}
\end{figure}

\section{Results}
\label{result}
\begin{figure}[htb!]
    \centering
    \includegraphics[height=3.0in,width=3.5in]{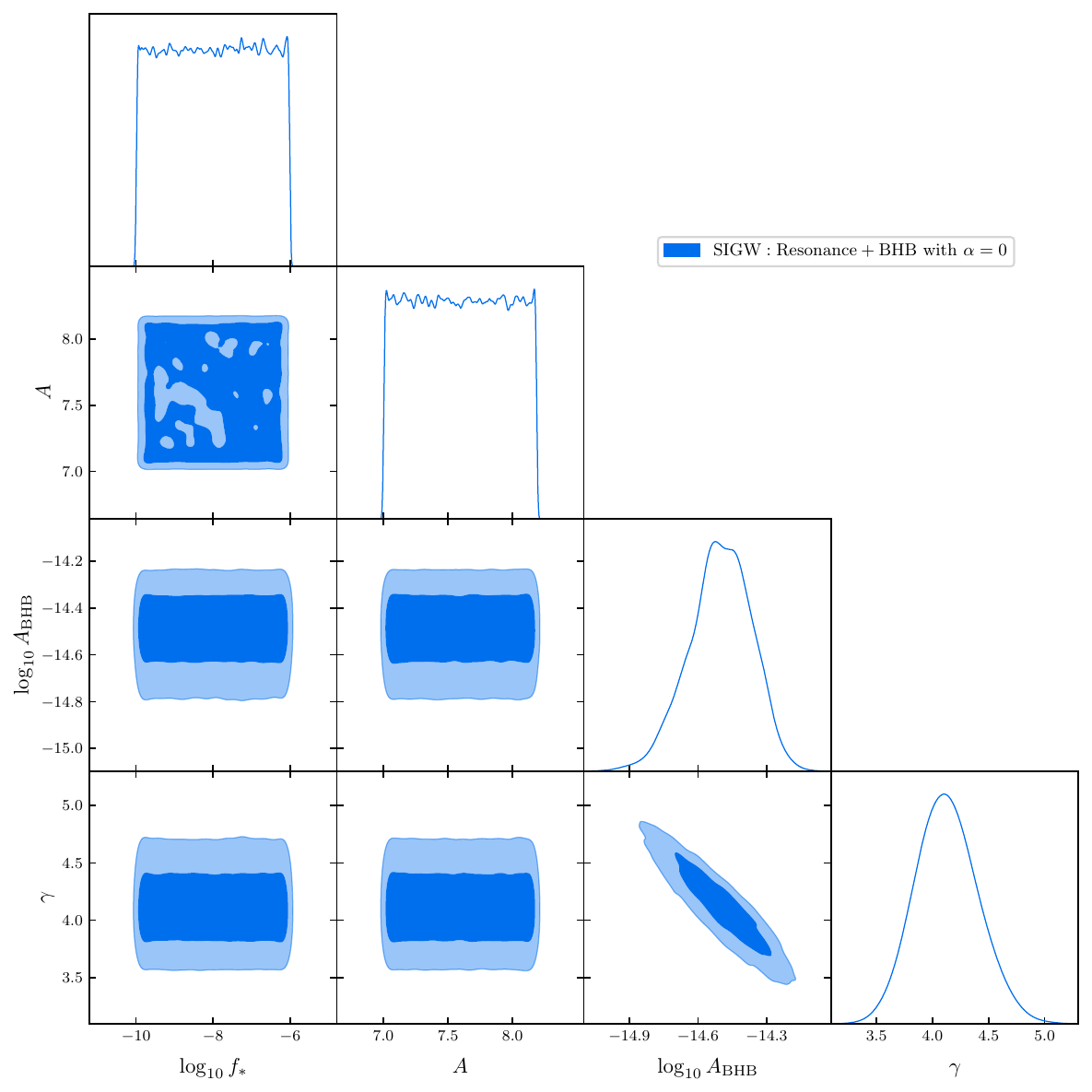}
    \caption{The current NANOGrav constraint on $A$ with the astrophysical stochastic background for $\alpha=0.0, \beta =0$ case. }
    \label{fig:corner_bhb_alpha0}
\end{figure}

\begin{figure}[htb!]
    \centering
    \includegraphics[height=3.0in,width=3.5in]{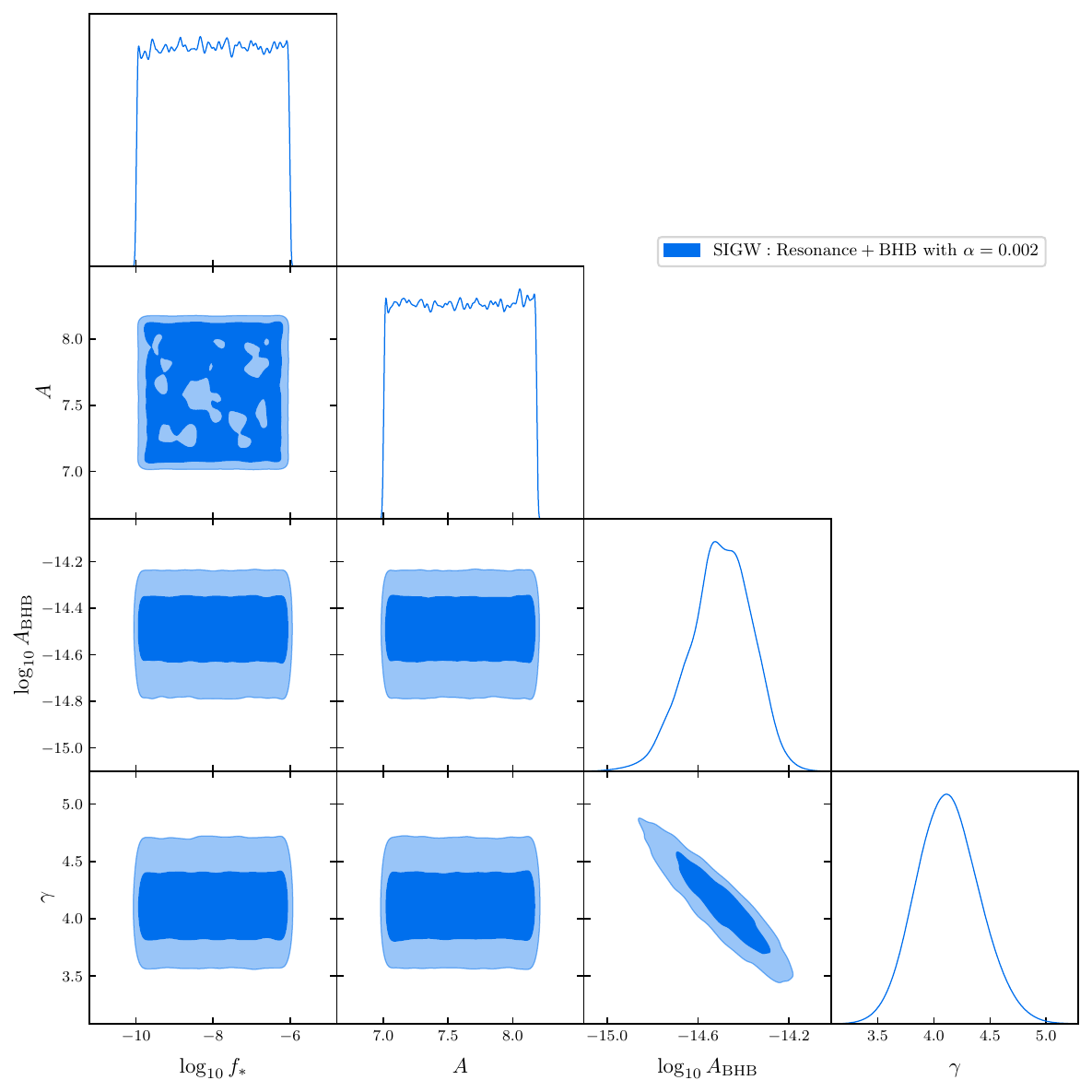}
    \caption{The current NANOGrav constraint on $A$ with the astrophysical stochastic background for fixed values of  $\alpha=0.002, \beta =0$ case.}
    \label{fig:corner_bhb_alpha0002}
\end{figure}
\begin{table}[htb!]
\centering
\begin{tabular}{| m{10em} | m{5cm} |}
\hline
Parameters & Prior\\
\hline
$A$ & $\text{Uniform}~[7.0, ~8.19]$\\[1ex]
\hline
$f_* ~/[\rm Hz]$  &  $\text{Uniform}~[10^{-10}, ~10^{-6}]$\\[1ex]
\hline
$N$ & $\text{Uniform}~[10, ~30]$\\[1ex]
\hline
$\lambda$ & $\text{Uniform}~[0.30, ~1.0]$\\[1ex]
\hline
$(\log_{10}A_{\rm BHB},~\gamma)$ & $\text{Normal}~[\boldsymbol{\mu}_{\rm BHB}, \boldsymbol{\sigma}_{\rm BHB}]$\\[1ex]
\hline
\end{tabular}
\caption{Prior distributions used for the parameter estimation with the case of fixed $(\alpha,~\beta) = (0, 0),(0.002, 0)$. $f_*$ is the peak frequency of the primordial scalar power spectrum.}
\label{table: param_prior_alpha0002}
\end{table}

\begin{table}[htb!]
\centering
\begin{tabular}{| m{15em} | m{3cm} |}
\hline
Model & Acceptance Ratio\\
\hline
$\text{Resonance}~( \alpha=0.00)$ & $~~~0.18$\\[1ex]
\hline
$\text{Resonance}~( \alpha=0.002)$  &  $~~~0.22$\\[1ex]
\hline
$\text{Resonance+SMBHB}~( \alpha=0.00)$ & $~~~0.52$\\[1ex]
\hline
$\text{Resonance+SMBHB}~( \alpha=0.002)$ & $~~~0.52$\\[1ex]
\hline
\end{tabular}
\caption{Acceptance ratio for the different models considered. }
\label{acceptance_table}
\end{table}

We have carried out the parameter estimation against our model with a public Python module PTArcade~\cite{Mitridate:2023oar} (see also \cite{Ellis:2023oxs,Bhaumik:2023wmw,Servant:2023mwt,Depta:2023qst,Gouttenoire:2023nzr}). The prior distributions for the parameter estimation are shown in Table~\ref{table: param_prior_alpha0002}. The parameters $\boldsymbol{\mu}_{\rm BHB}, \boldsymbol{\sigma}_{\rm BHB}$ give the mean and the covariance matrix
\begin{equation}
\boldsymbol{\mu}_{\rm BHB} = 
\begin{pmatrix}
-15.6 \\
4.7 \\
\end{pmatrix}, \\
\boldsymbol{\sigma}_{\rm BHB} = 
\begin{pmatrix}
0.28 & -0.0026 \\
-0.0026 & 0.12 \\
\end{pmatrix}
\end{equation}
of the two-dimensional normal distribution for the amplitude $\log_{10}A_{\rm BHB}$ and the spectral index $\gamma$ of the astrophysical stochastic background. The reconstructed posterior for the parameter $A$ without the astrophysical background is shown in Figs.~\ref{fig:corner_alpha0}, ~\ref{fig:corner_alpha0002} and the posterior with the astrophysical background is shown in Figs.~\ref{fig:corner_bhb_alpha0}, ~\ref{fig:corner_bhb_alpha0002}. The reconstructed posterior for the number of $e$-foldingn $N$ and the Yukawa coupling $\lambda$ is exhibited in Figs.~\ref{fig:corner_N_lambda_alpha0}, ~\ref{fig:corner_N_lambda_alpha0002}. 
The acceptance ratio is 0.18 for the parameter estimation of $A$ without the astrophysical background in the case of $(\alpha, ~\beta)= (0,0)$, 0.22 for $(\alpha, ~\beta)= (0.002,0)$, and 0.52 with the astrophysical background for $(\alpha, ~\beta)= (0,0),(0.002,0)$. For the $(N, \lambda)$ parameter search, the acceptance ratio is 0.10 and 0.11 with $(\alpha, ~\beta)=(0,0), (0.002,0)$ respectively \ref{acceptance_table}. 
The reconstructed primordial GW energy-density power spectra with a reference value are displayed in Fig.~ \ref{fig:Omggw_alpha0},  \ref{fig:Omggw_alpha0002}.

\section{Conclusion}
\label{conc}

Different Pulsar Timing Array  observations 
probe scales otherwise impossible to probe through laser interferometry techniques. The recent findings  by the PTA collaborationshowever,  has raised the question of how to understand the reported enhancement of the GW spectrum in the $n$Hz frequency range. Many proposals have been made to explain the observation. One of the best-suited explanations as quoted in the NANOGrav paper, is that of a sharp increment in the scalar power spectrum modeled by a Dirac delta function. This leds to an increment of the induced GW power spectrum  at  second order in perturbation theory and is dubbed as SIGW-DELTA. 

Although, statistically, SIGW-DELTA seems to be one of the best-suited models, justifying such an effect from theory is rather difficult. Interestingly, however, if there is a resonant production of a fermionic field that takes place during inflation as explained in section~\ref{resonance creation}, the effect will mimic the SIGW-DELTA. Thus, the explanation of the NANOGrav observation from the idea of resonance creation of the fermionic field becomes very relevant. 

As shown by our results, for a specific choice of the model parameters, one can obtain a good reproduction of the observation. Of course, adding the contribution from supermassive black-hole binaries(SMBHB) gives a better acceptance ratio as in the case of the NANOGrav collaboration findings, at least for the SIGW-DELTA case.

Nevertheless, there are a couple of remaining questions regarding the PBH overproduction and the effect of the variation of the equation of state ($\omega$) at the time of the enhancement of the GWs.  The authors plan to come back to this in a future work \cite{Domenech:2019quo,Liu:2023pau}.

There are also quite a few other avenues to explore, such as the effect of noncanonical inflation \cite{Bhattacharya:2018xlw}, warm inflation \cite{Gangopadhyay:2020bxn,Correa22}, and alternate gravity models such as Einstein Gauss Bonnett theory \cite{Pozdeeva:2020apf,Gangopadhyay:2022vgh,Khan:2022odn,Kawai:2021edk,Kawai:2021bye} that could lead to some interesting outcomes in this regard. The authors will also come back to this in the future. 

Although SIGW seems to be the best possible explanation for the observed spectrum by PTA collaborations, the natural consequence of the production of SIGW is the production of PBHs. There is a recent ongoing controversy \cite{Kristiano1,Kristiano2,Choudhury1,Choudhury2,Choudhury3,Choudhury4,Bhattacharya:2023ysp,Riotto:2023hoz,Riotto:2023gpm,Firouzjahi:2023ahg,Tada:2023rgp,Fumagalli:2023loc,Fumagalli:2023hpa} regarding the effect of loop contributions putting stringent bounds on the scales associated with the mass of the PBHs. It will be interesting to explore the idea proposed here in the light of such constraints.

\vspace{0.3cm}
\textit{Acknowledgments.}---
The work of M.R.G. is supported
by DST, Government of India under the Grant Agreement number IF18-PH-228, and by Science and Engineering Research Board (SERB), DST, Government of India under the Grant Agreement number CRG/2022/004120 (Core Research Grant). Work of G.J.M. at the University of Notre Dame supported by DOE nuclear theory grant DE-FG02-95-ER40934. 
T.K. is supported in part by the National Key R$\&$D Program of China (2022YFA1602401) and Grants-in-Aid for Scientific Research of Japan Society for the Promotion of Science (20K03958). Yogesh would like to thank Puneet Kumar Sharma for useful discussions. AM wants to thank IASc for providing the opportunity to work as a summer intern under MRG and SGT University for providing the required hospitality during the beginning of the work. R.I. is financially supported by JST SPRING, Grant Number JPMJSP2125, and would like to take this opportunity to thank the “Interdisciplinary Frontier Next-Generation Researcher Program of the Tokai Higher Education and Research System.”


\end{document}